# Optical Orientation of Excitons in CdSe Self-Assembled Quantum Dots

S. Mackowski*, T. Gurung, H. E. Jackson, and L. M. Smith

*Department of Physics University of Cincinnati, Cincinnati OH 45221-0011, USA*

J. K. Furdyna and M. Dobrowolska

*Department of Physics University of Notre Dame, Notre Dame IN, USA*

**Abstract**

We study spin dynamics of excitons confined in self-assembled CdSe quantum dots by means of optical orientation in magnetic field. At zero field the exciton emission from QDs populated via LO phonon-assisted absorption shows a circular polarization of 14%. The polarization degree of the excitonic emission increases dramatically when a magnetic field is applied. Using a simple model, we extract the exciton spin relaxation times of 100 ps and 2.2 ns in the absence and presence of magnetic field, respectively. With increasing temperature the polarization of the QD emission gradually decreases. Remarkably, the activation energy which describes this decay is independent of the external magnetic field, and, therefore, of the degeneracy of the exciton levels in QDs. This observation implies that the temperature-induced enhancement of the exciton spin relaxation is insensitive to the energy level degeneracy and can be attributed to the same excited state distribution.

* electronic mail: seb@physics.uc.edu

## I. INTRODUCTION

The methods of optical orientation and alignment of excitons [1] are widely used for studying the spin dynamics of carriers in semiconductor quantum structures. The principle of this technique is based on the observation that any losses of polarization observed in the photoluminescence (PL) emission are due to spin relaxation during the lifetime of the carriers [1]. Moreover, optical orientation methods can be significantly extended through the application of external magnetic fields. Applying the magnetic field parallel or perpendicular to the k-vector of the light splits the energy levels due to the Zeeman. Due to its versatility, optical orientation has been previously used not only to study the spin relaxation processes in semiconductors [2,3] but also to determine the effective g-factor anisotropy and its width dependence in semiconductor quantum wells [4].

In particular, optical orientation experiments performed on self-assembled quantum dots (QD), in addition to time-resolved PL measurements [5-7], have provided a means for determining the effect of the QD asymmetry and of external magnetic fields on exciton spin relaxation [8,9]. It has been found that for symmetric QDs at zero magnetic field, where the energy levels are degenerate, the exciton spin relaxation is extremely rapid, of the order of 10 ps [10]. In contrast, by removing the degeneracy -- either by the in-plane asymmetry of the QD potential [11] or by an external magnetic field [11] -- one can enhance the exciton spin relaxation time by orders of magnitude [10]. These experimental efforts have been accompanied by a number of elaborate theoretical models [12-14]. The major finding of these models is that the dominant mechanism responsible for the exciton spin relaxation in QDs is the mixing between different spin levels, mainly via spin-orbit interaction [12]. The actual spin-flip transitions between split energy levels are mediated by low-frequency acoustic phonons.

In this work we use optical orientation to study the spin relaxation of excitons confined to CdSe QDs grown by molecular beam epitaxy. In clear contrast to other II-VI semiconductor QDs, we observe a strong (14%) polarization of the PL emission even at B=0 T. With increasing magnetic field the polarization of the emission increases by a factor of three at B=3 T. Using simple estimates, we find that the exciton spin relaxation time is of the order of 100 ps and 2.2 ns in the absence and presence of an applied magnetic field, respectively. In addition, an activation energy of 8 meV has been calculated from the decrease of the PL polarization with increasing temperature, which might be plausibly associated with higher energy levels in these QDs. Importantly, the value of this activation energy does not depend on the magnetic field, which suggests the mechanism of the temperature induced polarization quenching is independent of the degeneracy of the exciton levels.

## II. SAMPLE AND EXPERIMENT

The self-assembled CdSe QD sample was fabricated by molecular beam epitaxy on a (100) GaAs substrate. Prior to deposition of the QD layer, a ZnSe buffer several micrometers thick was grown. The dots were formed by depositing 2.5 monolayers of CdSe at a temperature of 280 C [15]. This QD sample has been previously characterized by PL spectroscopy [16]. Non-resonantly excited PL emission is centered at E=2.235 eV, and features a Gaussian lineshape with the full width at a half maximum (FWHM) of 60 meV. Substantial shift of the QD emission to energies higher than the CdSe band gap (1.8 eV) indicates both strong spatial confinement of carriers localized by these QDs and some degree of alloying with the ZnSe matrix. On the other hand, the relatively large broadening of the PL emission line is ascribed to statistical fluctuations of QD parameters such as size, shape, strain distribution and chemical composition. From earlier single-dot PL measurements in magnetic fields the exciton effective g-factor in single CdSe QDs is known to be approximately –1 ± 0.2 [17].

Photoluminescence (PL) was measured with the sample mounted on a cold finger in a continuous flow helium cryostat. The temperature of the sample was varied between 6 K and 120 K. The results for the large QD ensemble were obtained with the laser focused to a spot of 30-μm diameter. The experiments on single CdSe QDs were carried out with the laser focused to a 1.7-μm diameter spot using a microscope objective (50X/0.5NA). In order to excite spin-polarized excitons directly into the ground states of the QDs, an Ar ion pumped Pyrromethane dye laser was tuned to 2.294 eV, i.e. within the non-resonantly excited emission band. A combination of Glan-Thomson linear polarizers with Babinet-Soleil compensators controlled and measured the polarization of the excitation and QD emission. Magnetic fields up to 3 T were applied in the Faraday configuration. Temperature-dependent and polarization-resolved experiments were performed at B=0, 1 and 2.5 T. The PL emission was dispersed through a triple DILOR spectrometer working in a double-subtractive mode and detected by a liquid nitrogen cooled CCD detector.

## III. OPTICAL ORIENTATION OF EXCITONS AT T=5K

### a. Ensemble of CdSe Quantum Dots

In Fig. 1 we show low temperature (T=6 K) resonantly excited PL spectra measured for CdSe QDs at B=0 T. The excitation laser at the energy of 2.294 eV is (a) $\sigma^+$ - polarized and (b) $\sigma^-$ - polarized, with both $\sigma^+$ - polarized (solid points) and $\sigma^-$ - polarized (open points) emissions displayed for each excitation. Because of the selection rules for excitonic optical transitions in QDs, this configuration of the excitation and detection polarizations at B=0T probes dots with degenerate exciton energy levels [8], characterized by angular momentum J=±1. Such degenerate exciton levels are seen only in symmetric QDs, in contrast to asymmetric QDs, where at zero magnetic field the states are split into linearly polarized doublets [8,11].

The resonantly excited PL spectra (see Fig. 1) of the CdSe QDs are strongly modified as compared to non-resonant PL [16]. Instead of a smooth Gaussian lineshape a complicated spectral response is observed, consisting of pronounced narrow lines (FWHM of several meV) superimposed on the inhomogeneously broadened non-resonantly excited emission [18]. These lines, spaced from the excitation laser by multiples of LO phonon energies, are attributed to the emission of QDs that are populated directly into their ground states through LO phonon-assisted absorption [18]. In this process, a photon creates a virtual state comprised of the exciton corresponding to the QD ground state and an integer number of LO phonons. As has been shown recently for CdTe QDs, LO phonon-assisted absorption provides a straightforward means for controlled excitation of spin-polarized excitons into the ground states of QDs [8].

As can be seen in Fig. 1 (a), when the ensemble of CdSe QDs is excited with $\sigma^+$ circularly polarized laser, the emission at zero magnetic field is also $\sigma^+$ circularly polarized. Furthermore, PL spectra obtained at B=0 T for $\sigma^-$ - polarized excitation (see Fig. 1 (b)) are also predominantly $\sigma^-$ - polarized. This result is in clear contrast to previous studies, where for CdTe QDs [8] and InAs QDs [5] no circular polarization has been observed at zero magnetic fields. The absence of polarization in these QD systems has been attributed to very rapid spin relaxation of the excitons confined to symmetric QDs [8] or, alternatively, to the random distribution of the exchange splitting within the QD ensemble [5]. Therefore, the results shown in Fig. 1 implies that (1) there are symmetric QDs in this sample, and (2) a significant number of excitons in symmetric CdSe QDs maintain the spin polarization during the exciton recombination time (measured to be 700ps [16]).

By applying an external magnetic field to the symmetric QDs we remove the degeneracy of the exciton energy levels via the Zeeman interaction [11]. Importantly, the negative sign of the effective g-factor of excitons in the CdSe QDs [8] implies that the emission of the lower exciton level at B>0 T is always $\sigma^-$- polarized. In Fig. 2 we display the resonantly excited PL spectra measured at B=2 T for both (a) $\sigma^+$ and (b) $\sigma^-$ circularly polarized excitations. In each case we observe both the $\sigma^+$ - polarized (solid points) and $\sigma^-$ - polarized (open points) emissions. In an applied magnetic field the polarization of the emission of QDs populated via LO phonon-assisted absorption increases substantially, presumably due to an increase of the exciton spin relaxation time for the non-degenerate exciton levels [5]. Moreover, in contrast to the results at B=0 T, the absolute value of the polarization of the LO phonon replicas at B=2 T depends on the polarization of the excitation. Namely, we observe larger polarization of the PL emission for $\sigma^-$ - polarized excitation (see Fig. 2 (a) and (b) and the discussion below). Qualitatively similar behavior, previously observed for CdTe QDs [8], has been ascribed to partial thermalization of carriers between exciton levels split by the external magnetic field.

**b. Single CdSe Quantum Dots**

It should be noted that the PL spectra displayed in Fig. 1 have been measured for a large ensemble of CdSe QDs, so that they represent statistically averaged behavior of the PL polarization. One could therefore expect that the polarization of single QD emissions might show some variation between different QDs. In Fig. 3 we present the PL spectra detected on two single CdSe QDs at B=0 T. The resonant laser excitation was $\sigma^+$ - polarized, and emissions in both circular polarizations were analyzed. We insured that both of these QDs were excited via LO phonon-assisted absorption. As can be seen, in the case of QD#1 (see Fig. 3 (a)) the emission is strongly polarized, and the polarization is in the same direction as the polarization of the excitation. In sharp contrast, the QD#2 (see Fig. 3 (b)), shows virtually no polarization at B=0 T. Since the polarization of the PL emission can be directly related to the exciton spin relaxation in QDs, the observed variation of the polarization suggests that the spin dynamics in QDs is very sensitive to the specific parameters of each individual QD.

**c. Analysis of the polarization of the CdSe quantum dot ensemble at T=5K**

In order to extract the PL polarization of the QDs populated through LO phonon-assisted absorption we fit the spectra shown in Fig. 1 with three Gaussian lineshapes. The comparison between the PL spectra fitted and measured at B=0 T and B=2 T is given in Figs. 4 (a) and 4 (b), respectively. The two narrower Gaussians (dashed lines) represent the QDs populated via LO phonon - assisted absorption, while the third, much broader fit (dotted line), is related to direct excitations between the excited states and the ground states [8,18]. In this way we obtain reasonable agreement between the experiment and the fitted spectra. Since we assume that the energy distributions of the excited and ground states in the QD ensemble are identical, the width and the energy of the broader Gaussian is given by the non-resonantly excited PL [18]. Also, since we have no information about the energy structure of the excited states we will assume that the excitons in QDs populated through excited state excitations are unpolarized.

The intensities of the LO phonon replicas have been extracted from the fits as a function of polarization of the excitation and of the applied magnetic field. In Fig. 5 (a) we plot the polarization of the first LO phonon replica as a function of magnetic field for both $\sigma^+$ (solid squares) and $\sigma^-$ (solid circles) circularly polarized excitations. The magnitude of the polarization is defined as $P=(I^+ \; I^-)/(I^+ + I^-)$, where $I^+$ and $I^-$ are the intensities of $\sigma^+$ and $\sigma^-$ emissions, respectively. Remarkably, we find that the first LO phonon replica features a substantial (14%) polarization at B=0 T. Furthermore, this polarization shows a pronounced increase with applied magnetic field (three-fold at 3 T). The simple analysis described in Ref. **2** yields a value of the exciton spin relaxation time to be about 100 ps for symmetric CdSe QDs at B=0 T. We note that this value is an order of magnitude larger than the exciton spin relaxation time at zero magnetic fields estimated for CdTe QDs [8], where no polarization of the PL emission has been observed without magnetic field.

Furthermore, as discussed in detail in Ref. 10, by fitting the magnetic field dependence of the difference between polarizations (ΔP) measured for $\sigma^-$ and $\sigma^+$ - polarized excitations, one can estimate the exciton spin relaxation time for QDs with non-degenerate spin levels. In Fig. 5 (b) we plot the difference ΔP (open points) obtained for the emission of the CdSe QDs as a function of the external magnetic field. The solid line represents the fit to the experimental data using the expression [10]:

$$\Delta P = \frac{2\tau_R (e^{\Delta E / k_B T} - 1)}{\tau_R + e^{\Delta E / k_B T} (\tau_R + \tau_S)}$$

Here, $\tau_R$ and $\tau_S$ are the exciton recombination time and exciton spin relaxation time, respectively. The value of $\tau_R$ has been measured independently and it is equal to 700 ps [16]. The data is plotted for T=6 K and $\Delta E = g\mu_B B$ determined by assuming the effective g-factor –1. Using this relatively simple model, which does not include the effect of asymmetric QDs, we estimate that the exciton spin relaxation time in applied magnetic fields (where the exciton states are not degenerate) is 2.2 ns. This value is in agreement with estimates of $\tau_S$ in self-assembled QD based on other semiconductor systems [5,8].

To summarize, we have studied the exciton spin relaxation time in CdSe QDs by low temperature resonant PL measurements using circularly polarized. We find that, while at B=0 T most of the excitons in QDs populated through LO phonon-assisted absorption randomize their spins within the exciton recombination time, a significant fraction (14%) of the QDs maintains their initial spin polarization. On the other hand, in magnetic field we observe a dramatic increase of the exciton spin relaxation time, which we ascribe to the removal of the spin degeneracy of Zeeman-split exciton levels. The measured exciton spin relaxation time is 2.2 ± 0.3 ns. It is interesting to point out that this value of $\tau_S$ is very similar to the one measured for annealed CdTe QDs, which are characterized by similar QD sizes as the sample studied in this work [19]. This observation might possibly indicate that the exciton spin relaxation time decreases for QDs with larger sizes. Such behavior can be ascribed to the fact that for larger QDs the energy separation between ground states and excited states (e.g., states involving light holes) is reduced, which tends to increase the mixing between states of different spin, thus making spin scattering processes much more efficient.

## IV. TEMPERATURE DEPENDENCE OF THE OPTICAL ORIENTATION

We next consider the temperature dependence of the polarization of resonantly excited PL of CdSe QDs. In Fig. 6 we present resonantly excited PL spectra measured at T=70 K for the CdSe QD sample. The measurements were performed at (a) B=1 T and (b) B=2.5 T. In both cases the excitation was $\sigma^+$ -polarized, and the figure displays both $\sigma^+$ - (solid squares) and $\sigma^-$ - polarized (solid circles) emissions. With increasing temperature of the sample we observe a strong decrease of the polarization of the QD emission excited through the LO phonon-assisted absorption. Notably, at a given temperature larger polarization is obtained at the larger magnetic field of 2.5 T. The observed decrease of the PL polarization with increasing temperature is attributed to increased population of the acoustic phonons which are believed to dominate the spin flip processes of the QD excitons [5,12]. From the temperature dependence of the polarization we might also be able to determine the nature of electronic levels involved in these spin-flip transitions.

We analyze the temperature behavior of the polarization for different values of the external magnetic field. This allows us to extract the activation energy $E_A$, which governs the observed decay of the polarization by assuming a simple single exponential process given by the expression:

$$P(T) = P_0/(1+C\exp(-E_A/kT)),$$

where T is temperature, k is a Boltzmann constant and C and $P_0$ are constants. The result of the fits (solid lines) together with the experimental data (solid points) are plotted in Fig. 7 for B=0 T, B=1 T and B=2.5 T. It is rather remarkable that the value of the activation energy shows no dependence on the magnetic field, and is equal to 8±0.8 meV. Also, this value is smaller than activation energies obtained for InAs QDs, where the energies of 30 meV [5] have been observed. Such low activation energy could possibly be attributed to the excited states in these relatively large CdSe QDs (~8-10 nm) involving, for example, light-hole exciton states. Another possible origin for such states may come from the two-dimensional platelets that serve as precursors for the QD formation in this materials system [15].

## V. SUMMARY

In summary, we have used resonantly excited polarized PL spectroscopy to probe exciton spin relaxation processes in CdSe QDs. Our most important finding is to show that at zero magnetic fields and at low temperature we observe a 14% PL polarization of the CdSe QDs populated through LO phonon-assisted absorption. We attribute this strong optical orientation to the relatively slow relaxation (~100 ps) of the exciton spins in symmetric QDs. In an applied magnetic field, where the degeneracy of the exciton energy levels in QDs is lifted, the polarization of the emission increases by a factor of 3 in a field of 3T. The resulting exciton spin relaxation time is estimated to be 2.2 ns. In addition, we have also demonstrated that the activation energy of 8meV, extracted from the temperature dependence of the polarization, might possibly be associated with excited states in these QDs. Remarkably, the temperature induced suppression of the PL polarization is identical for all magnetic fields, which implies that both at B=0 T and B=2.5 T the same states in a given QD contribute to the spin relaxation process.

**ACKNOWLEDGEMENTS**

We acknowledge the support of NSF through Grants DMR 0071797, 0216374 and 0245227 and DARPA SpinS Program.

[1] *Optical Orientation,* edited by F. Meier and B. P. Zakharchenya, Modern Problems in Condensed Matter Sciences Vol. 8 (North-Holland, Amsterdam, 1984).

[2] R. I. Dzhioev, V. L. Korenev, B. P. Zakharchenya, D. Gammon, A. S. Bracker, J. G. Tischler, and D. S. Katzer, Phys. Rev. **B 66**, 153409 (2002).

[3] M. Potemski, J. C. Maan, A. Fasolino, K. Ploog, and G. Weimann, Phys. Rev. Lett., **63**, 2409 (1989).

[4] M. J. Snelling, G. P. Flinn, A. S. Plaut, R. T. Harley, A. C. Tropper, R. Eccleston, and C. C. Phillips, Phys. Rev. **B 44**, 11 345 (1991); E. Blackwood, M.J. Snelling, R.T. Harley, S.R. Andrews, C.T.B. Foxon, Phys. Rev. **B 50**, 14246 (1994).

[5] M. Paillard, X. Marie, P. Renucci, T. Amand, A. Jbeli, and J.M. Gerard, Phys. Rev. Lett. **86**, 1634 (2001).

[6] M. Scheibner, G. Bacher, S. Weber, A. Forchel, Th. Passow, and D. Hommel, Phys. Rev. **B 67**, 153302 (2003).

[7] T. Flissikowski, A. Hundt, M. Lowisch, M. Rabe, and F. Henneberger, Phys. Rev. Lett. **86**, 3172 (2001).

[8] S. Mackowski, T.A. Nguyen, T. Gurung, K. Hewaparkarama, H.E. Jackson, L.M. Smith, J. Wrobel, K. Fronc, J. Kossut, and G. Karczewski, Phys. Rev. B – in print 2004.

[9] R.I. Dzhioev, B.P. Zakharchenya, E.L. Ivchenko, V.L. Korenev, Yu.G. Kusraev, N.N. Ledentsov, V.M. Ustinov, A.E. Zhukov, and A.F. Tsatsul’nikov, Fiz. Tverd. Tela (St. Petersburg) **40**, 858 (1998) (Phys. Solid State **40**, 790 (1998)).

[10] S. Mackowski, T. A. Nguyen, H. E. Jackson, L. M. Smith, G. Karczewski, and J. Kossut, Appl. Phys. Lett. **83**, 5524 (2003).

[11] V.D. Kulakovskii, G. Bacher, R. Weigand, T. Kummell, A. Forchel, E. Borovitskaya, K. Leonardi, D. Hommel, Phys. Rev. Lett. **82**, 1780 (1999).

[12] A.V. Khaetskii and Y. V. Nazarov, Phys. Rev. **B 61**, 12639 (2000).

[13] E. Tsitsishvili, R. v. Baltz, and H. Kalt, Phys. Rev. **B 66**, 161405 (2002).

[14] L. M. Woods, T. L. Reinecke, and Y. Lyanda-Geller, Phys. Rev. B **66**, 161318 (2002).

[15] C.S. Kim, M. Kim, S. Lee, J.K. Furdyna, M. Dobrowolska, H. Rho, L.M. Smith, H.E. Jackson, E.M. James, Y. Xin, and N.D. Browning, Phys. Rev. Lett. **85**, 1124 (2000).

[16] S. Mackowski, S. Lee, J.K. Furdyna, M. Dobrowolska, G. Prechtl, W. Heiss, J. Kossut, and G. Karczewski, physica status solidi **b 229**, 469 (2002).

[17] K. P. Hewaparakrama, S. Mackowski, L. M. Smith, et al. - unpublished

[18] T. A. Nguyen, S. Mackowski, H.E. Jackson, L. M. Smith, J. Wrobel, K. Fronc, G. Karczewski, J. Kossut, M. Dobrowolska, J. Furdyna, and W. Heiss, Phys. Rev. B **70**, 125306 (2004)

[19] S. Mackowski, T. Gurung, H. E. Jackson, L. M. Smith, W. Heiss, J. Kossut, and G. Karczewski, unpublished, condmat/0409507

Figure captions:

**Figure 1.** Low temperature (T=6K) resonantly excited PL spectra of CdSe QDs measured at B=0T for (a) $\sigma^+$ and (b) $\sigma^-$ polarized excitations. Both $\sigma^+$ (solid points) and $\sigma^-$ (open points) polarizations are shown.

**Figure 2.** Low temperature (T=6K) resonantly excited PL spectra of CdSe QDs measured at B=2T for (a) $\sigma^+$ and (b) $\sigma^-$ polarized excitations. Both $\sigma^+$ (solid points) and $\sigma^-$ (open points) polarizations are shown.

**Figure 3.** Photoluminescence spectra of two single CdSe QDs obtained at B=0T showing (a) strongly polarized QD emission and (b) approximately unpolarized QD emission. The excitation was $\sigma^+$ polarized.

**Figure 4.** Comparison between experimentally measured (open points) and fitted (solid lines) PL spectra of CdSe QDs. The results obtained for $\sigma^+$ polarized excitation and detection at (a) B=0T and (b) B=2T are shown. Dashed lines represent LO phonon-assisted absorption while the dotted curve depicts excitations via excited states.

**Figure 5.** (a) Polarization of the CdSe QD emission corresponding to the first LO phonon replica obtained for $\sigma^+$ (solid points) and $\sigma^-$ (open points) polarized excitations versus external magnetic field. (b) Difference between polarizations shown in (a) together with calculated dependence (dashed line).

**Figure 6.** Resonantly excited PL spectra of CdSe QDs measured at (a) B=1T and (b) B=2.5T for $\sigma^+$ polarized excitations. The sample temperature is 70K. Both $\sigma^+$ (solid points) and $\sigma^-$ (open points) polarizations are shown.

**Figure 7.** Temperature dependence of the polarization of the CdSe QD emission measured at (a) B=0T, (b) B=1T, and (c) B=2.5T. Solid points represent the data, while lines are fitted assuming single activation energy process.

Figure 1

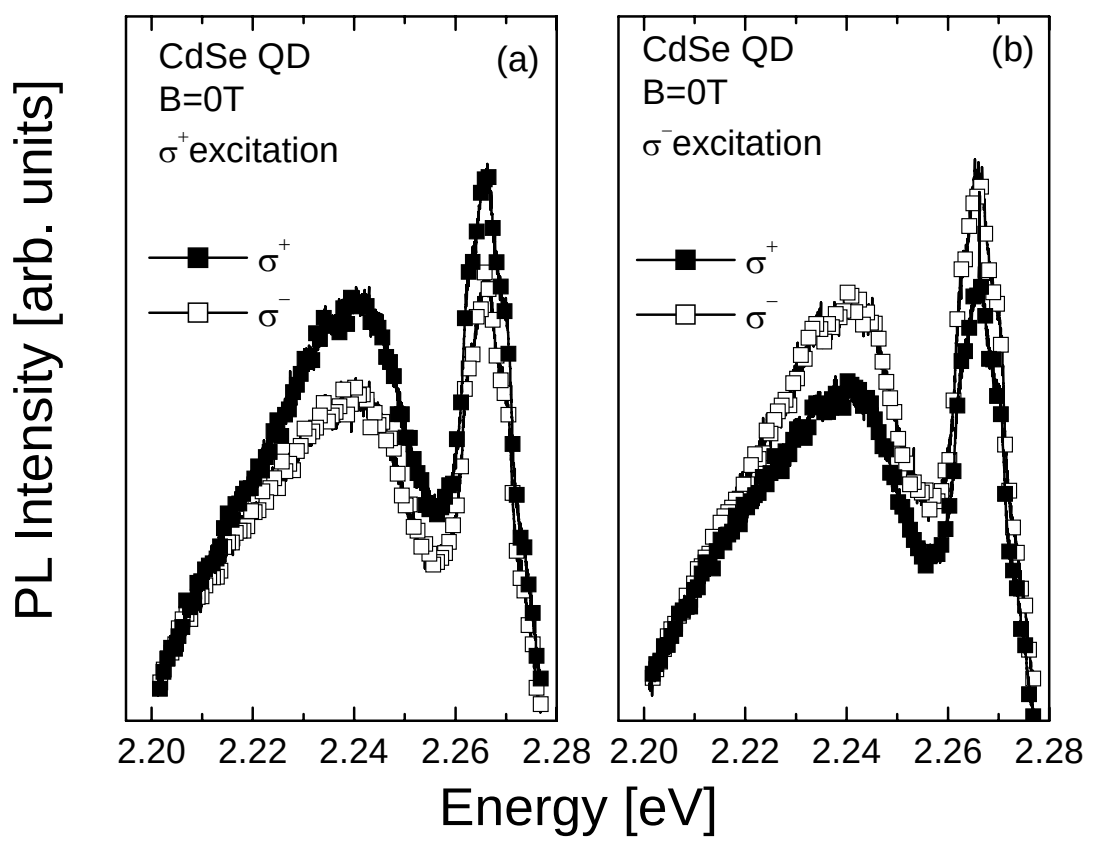

Figure 2

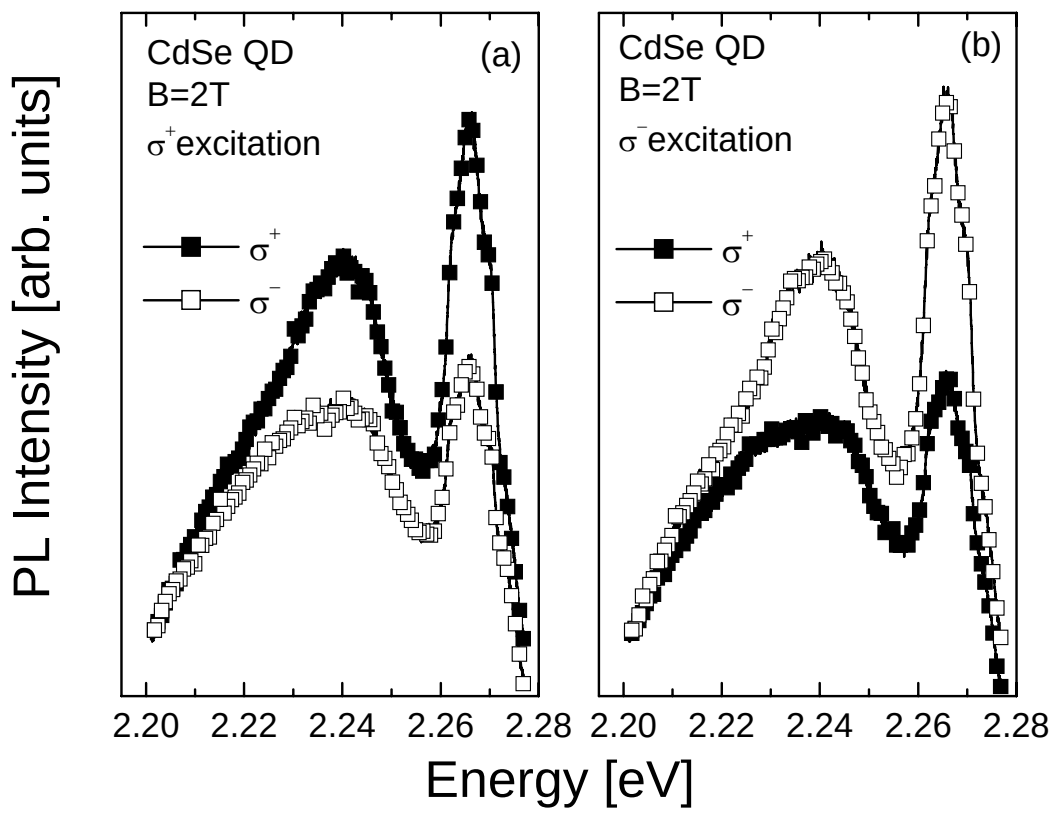

Figure 3

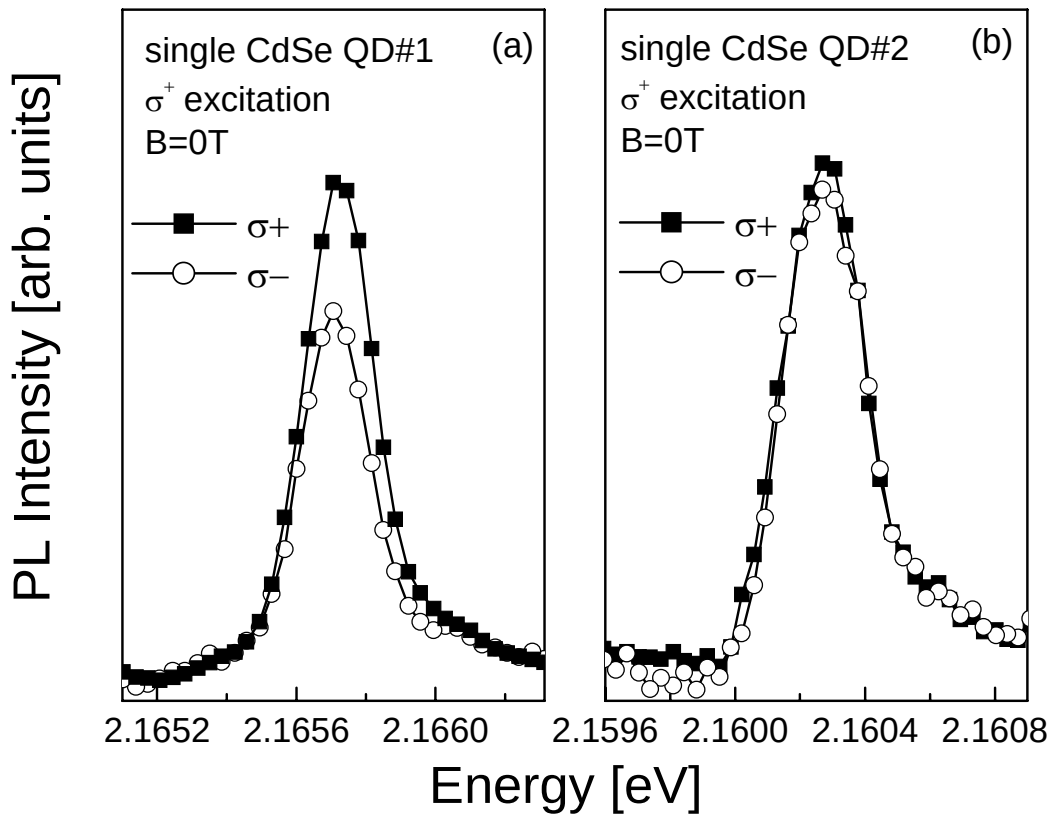

Figure 4

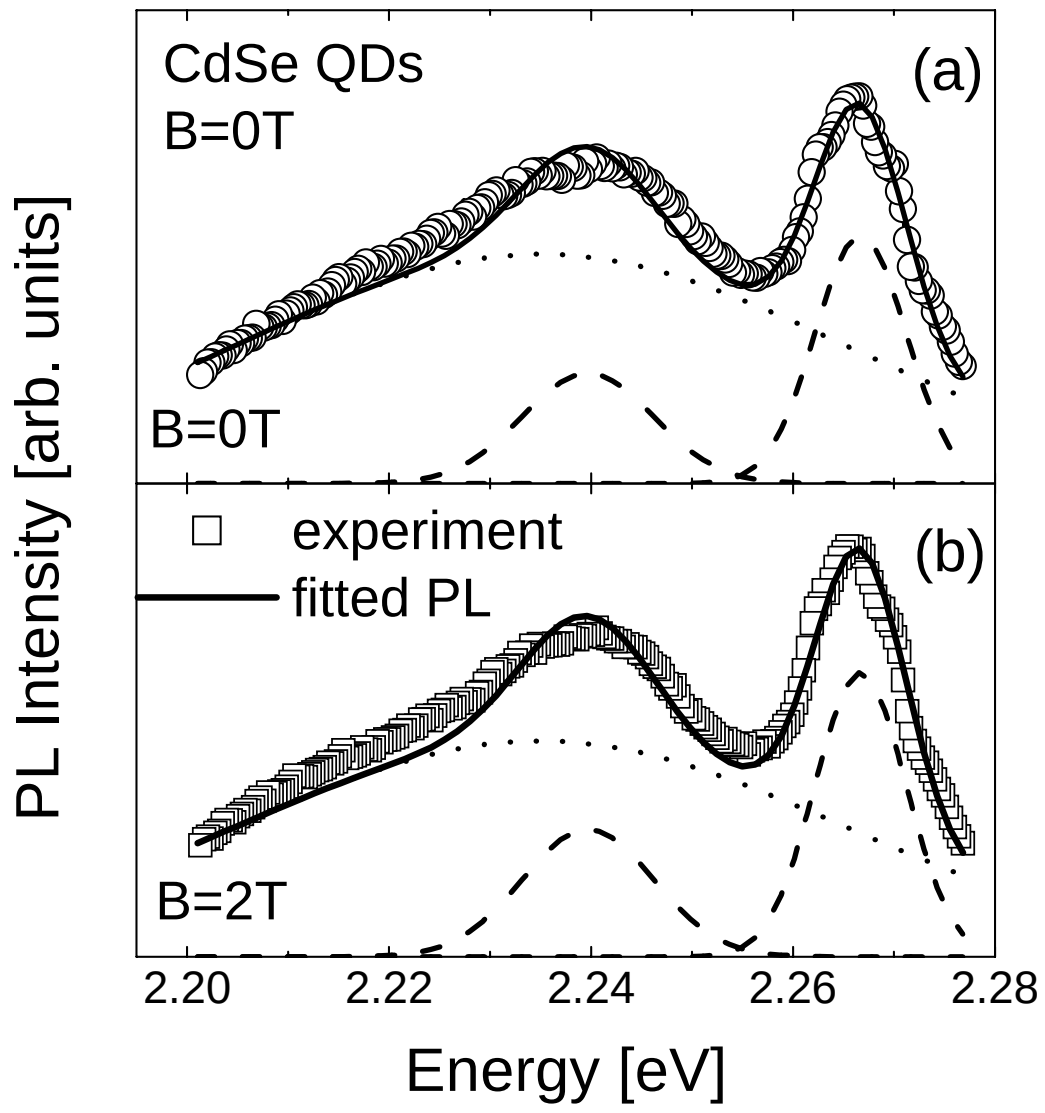

Figure 5

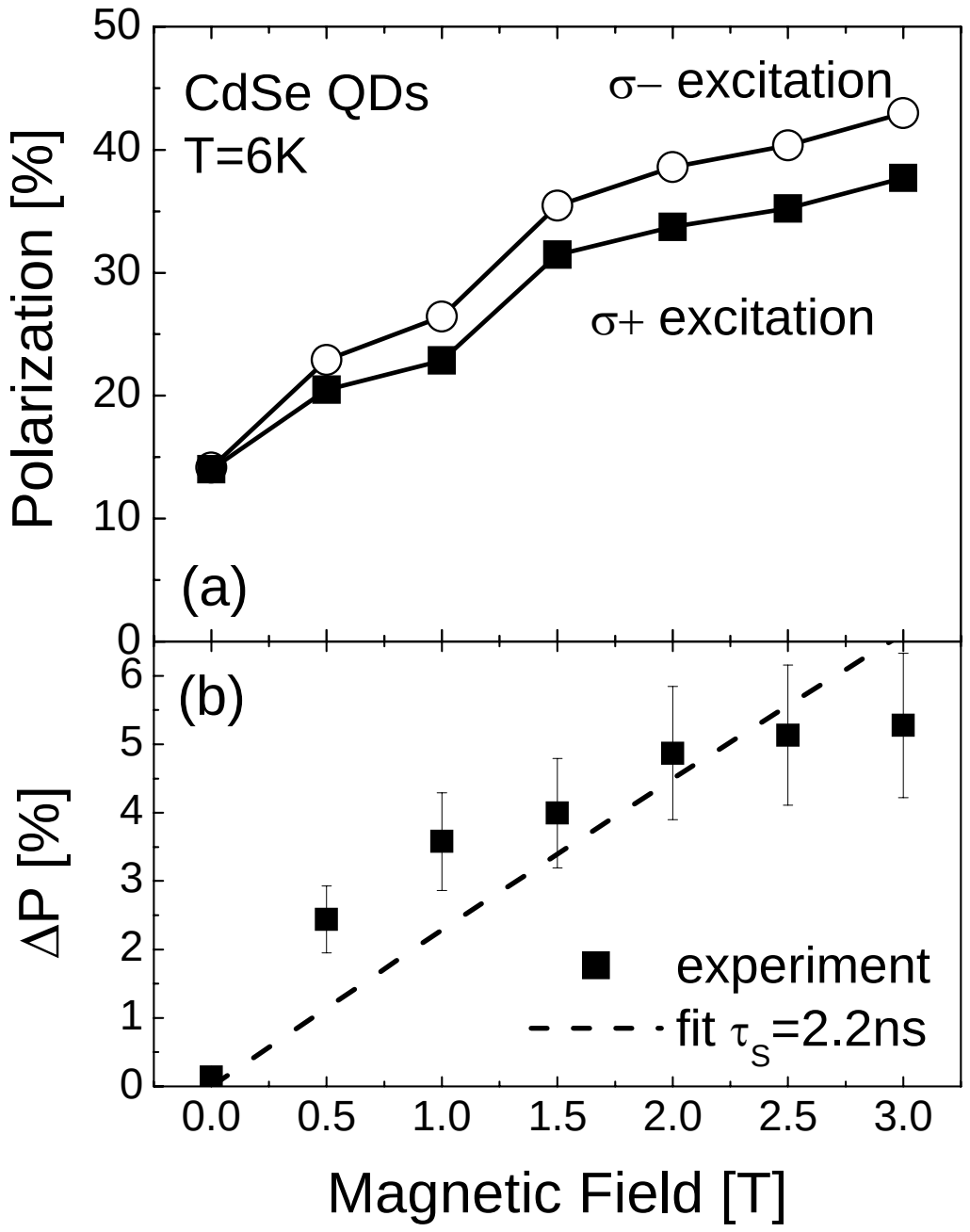

Figure 6

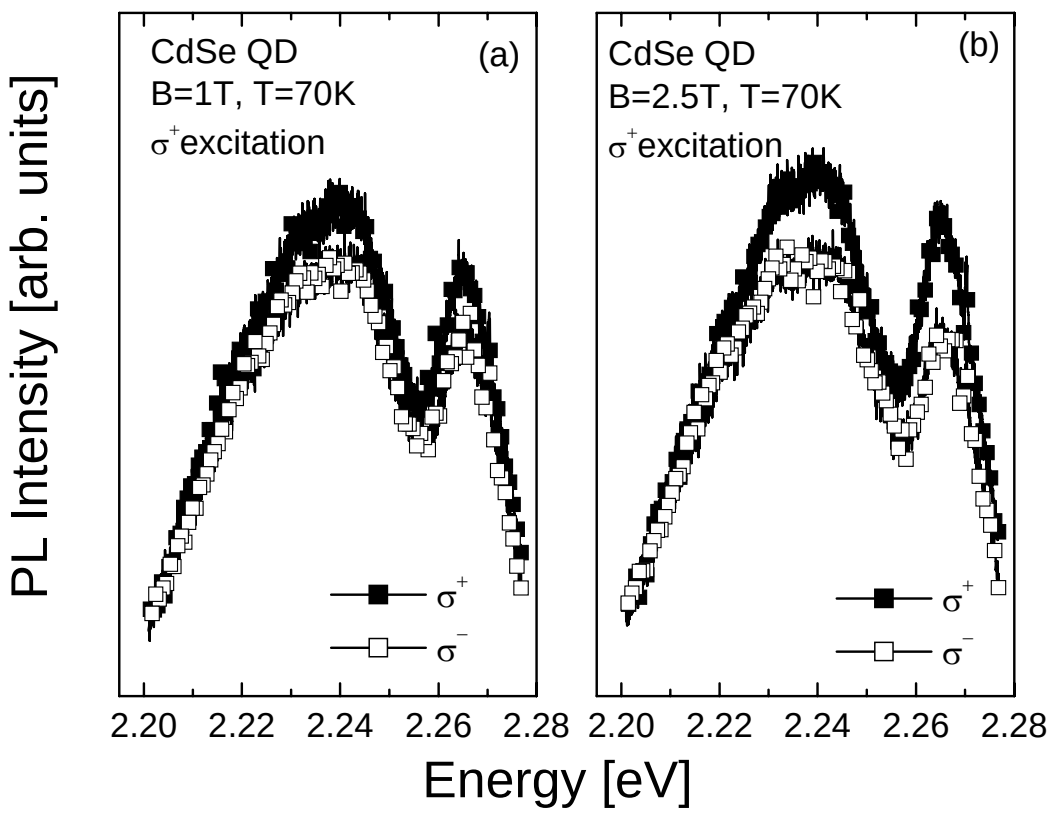

Figure 7

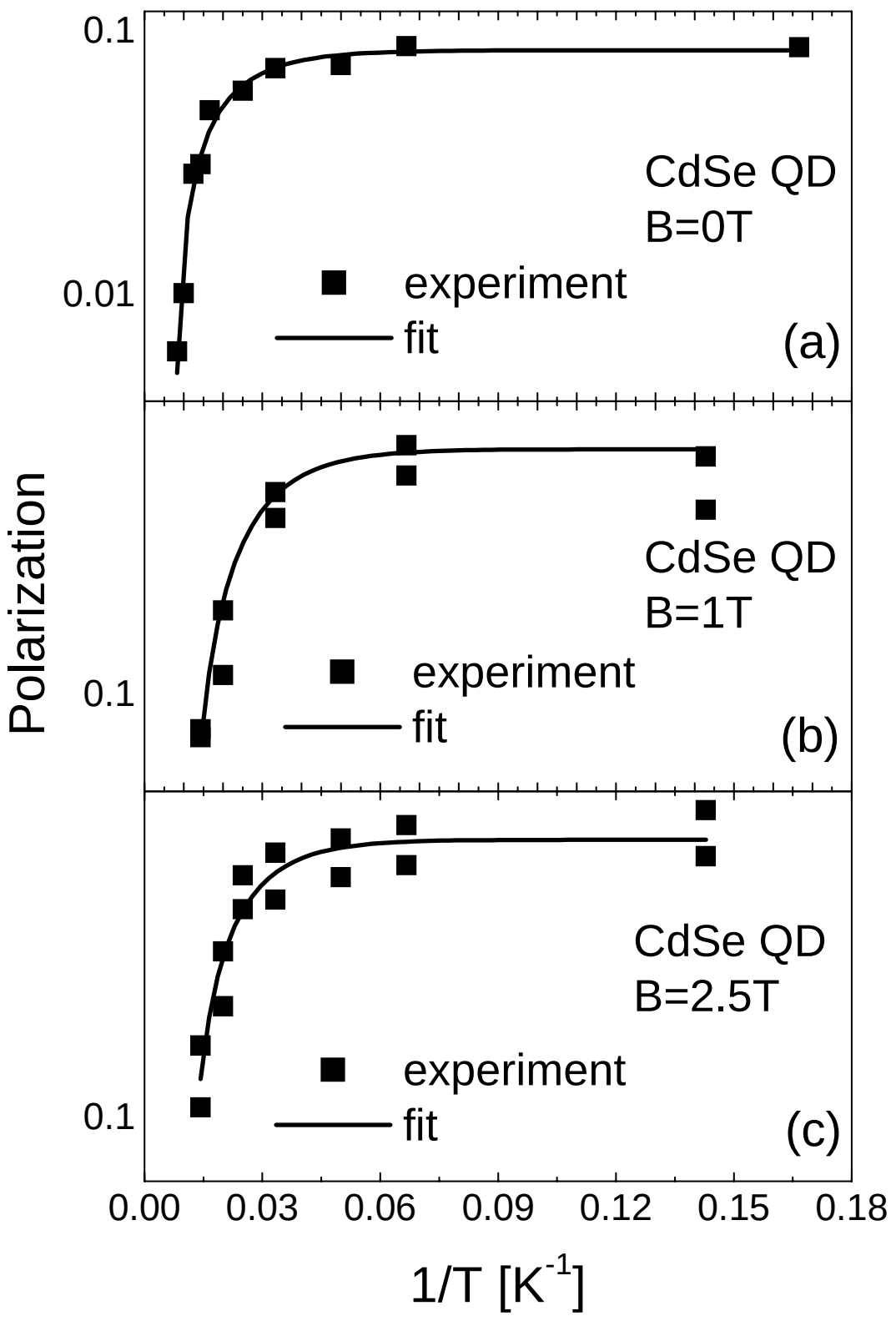